\documentclass[12pt]{iopart}
\usepackage{iopams}
\usepackage{epsfig}

\begin{document}

\tolerance = 10000

\title{Thermodynamics of SU(2) bosons in one dimension}
\author{Shi-Jian Gu, You-Quan Li, Zu-Jian Ying and Xue-An Zhao}
\address{Zhejiang Institute of Modern Physics,
 Zhejiang University, Hangzhou 310027, P.R. China}
\date{\today}

\begin{abstract}
On the basis of Bethe ansatz solution of two-component bosons with
SU(2) symmetry and $\delta$-function interaction in one
dimension, we study the thermodynamics of the system at finite
temperature by using the strategy of thermodynamic Bethe ansatz
(TBA). It is shown that the ground state is an isospin
``ferromagnetic" state by the method of TBA, and at high
temperature the magnetic property is dominated by Curie's law. We
obtain the exact result of specific heat and entropy in strong
coupling limit which scales like $T$ at low temperature. While in
weak coupling limit, it is found there is still no Bose-Einstein
Condensation (BEC) in such 1D system.
\end{abstract}

\maketitle

A two-component Bose gas has been produced in magnetically trapped
$^{87}Rb$ by rotating the two hyperfine states into each other
with the help of slightly detuned Rabi oscillation field
\cite{JEWilliams99}. It was noticed \cite{TLHo98} that the
properties of such Bose system can be different from the
traditional scalar Bose system once it acquires internal degree of
freedom. Bethe ansatz solution of SU(2) two-component bosons in
one dimension was obtained\cite{PPKulish81,YQLi01}. It was pointed out that
the ground state of such a system is an isospin ``ferromagnetic"
state\cite{YQLi01} which differs from that of spin-1/2 fermions in one
dimension, such as SU(2) Hubbard model\cite{EHLieb68} etc..

An interacting SU(2) boson field trapped in a one dimensional
ring of length $L$ can be modeled by the following Hamiltonian
\begin{equation}
H=\int dx \left [
\sum_a\partial_x\psi_a^*
\partial_x\psi_a+\frac{c}{2}\sum_{a, b}\psi_a^*\psi_a
\psi_b^* \psi_b \right ]
\label{eq:HAMILTONION}
\end{equation}
where $a, b=1, 2$ denotes the $z$-component of isospin. The Bethe
ansatz equations (BAE) of eq. (\ref{eq:HAMILTONION}) are obtained
as follows
\begin{eqnarray}
e^{ik_jL}=&&-\prod^{N}_{l=1}\frac{k_j-k_l+ic}{k_j-k_l-ic}
   \prod_{\nu=1}^{M}\frac{k_j-\lambda_\nu-ic/2}{k_j-\lambda_\nu+ic/2}
    \nonumber\\
1=&&-\prod^{N}_{l=1}
    \frac{\lambda_\gamma-k_l-ic/2}{\lambda_\gamma-k_l+ic/2}
     \prod^{M}_{\nu=1}
      \frac{\lambda_\gamma-\lambda_\nu+ic}{\lambda_\gamma-\lambda_\nu-ic}
\label{eq:BAE}
\end{eqnarray}
where $M$ denotes the total number of down isospins. Eq.
(\ref{eq:BAE}) differs from the BAE of scalar bosons with periodic
condition\cite{EHLieb63}. The second equation of eqs.
(\ref{eq:BAE}) of isospin rapidity $\lambda$ arises from the
application of quantum inverse method\cite{LDFaddeev84}, which
can be inferred from spin-1/2 fermions\cite{CNYang67} too.
However, the symmetry of bosons wave function leaves the first
term on the right side of first equation of eqs. (\ref{eq:BAE}),
which does not appear in the BAE of fermions.

The strategy we use here is the thermodynamic Bethe ansatz (TBA)
which was pioneered by C. N. Yang and C. P. Yang for the case of
the delta-function Bose gas\cite{CNYang69}. It is used to derive
a set of nonlinear integral equations called TBA equations, which
describe the thermodynamics of the model at finite temperature.
Moreover, the $\lambda$s can be complex roots which should form a
``bound state" with other $\lambda$s\cite{MTakahashi72} when
$T\neq 0$, which arises from the consistency of both sides of the
BAE.  For ideal $\lambda$ strings of length $m$ the rapidities are
$\Lambda_a^{nj}=\lambda_a^n+(n+1-2j)iu+O(\exp(-\delta N))$. Here
$u=c/2$, $a$ enumerates the strings of the same length $m$, and
$j=1, \dots, m$ counts the $\lambda$s involved in the $a$th
$\lambda$ string of the length $m$, $\lambda_a^n$ is the real part
of the string.

Taking logarithm of the BAE (\ref{eq:BAE}) by using string
hypothesis we arrive at the following discrete Bethe ansatz
equations
\begin{eqnarray}\label{eq:dBAE}
2\pi I_j&&=k_j L
    +2\sum_l\Theta_2(k_j-k_l)
   -\sum_{an}\Theta_n(k_j-\lambda_a^n)
 \nonumber \\
2\pi J_a^n&&=2\sum_l\Theta_n(\lambda_a^n-k_l)
   -2\sum_{bl,t\neq 0}A_{nlt}\Theta_t(\lambda_a^n-\lambda_b^l)
\label{eq:IJBAE}
\end{eqnarray}
where $\Theta_n(x)=\tan^{-1}(x/nu)$ and
\[
A_{nlt}=\left\{\begin{array}{ll}
1, & {\rm for}\; t=n+l, |n-l|\\
2, & {\rm for}\; t=n+l-2,\cdots,|n-l|+2\\
0,  & {\rm  otherwise.}
\end{array}\right.
\]
$I_j$ and $J_a^n$ play the role of quantum numbers for charge
rapidity and isospin rapidity respectively. In order to guarantee
linearly independent of wave function, all quantum number within
a given set of $\{I\}$ as well as that in $\{J\}$ should be
different. An arbitrary quantum number may be in a set (charge
roots) or missing (charge hole). In thermodynamic limit, the
distribution of charge rapidities becomes dense and it is useful
to introduce the density function for charge roots and holes
respectively. We denote with $\rho(k)$ and $\rho^h(k)$ the
density function of charge roots and holes, in a similar way,
with $\sigma_n(\lambda)$ and $\sigma_n^h$ the density function of
n-strings roots and holes on real axis. They are defined by
\begin{eqnarray}
\rho(k)+\rho^h(k)=&&(1/L)dI(k)/dk \nonumber \\
\sigma_n(\lambda)+\sigma_n^h(\lambda)=&&(1/L)
  dJ^n(\lambda)/d\lambda.
\end{eqnarray}
Then from eqs. (\ref{eq:dBAE}) we obtain a set of coupled integral
equations.
\begin{eqnarray}
\rho+\rho^h=&&\frac{1}{2\pi} + K_2(k)*\rho(k) -\sum_n
K_n(k)*\sigma_n(k)
\nonumber \\
\sigma_n+\sigma_n^h=&&K_n(\lambda)*\rho(\lambda)
 -\sum_{l,t\neq 0}A_{nlt}K_t(\lambda)*
 \sigma_l(\lambda)
\label{eq:DENSITY}
\end{eqnarray}
where $K_n(x)=nu/\pi(n^2u^2+x^2)$, and $*$ denotes a convolution.

In terms of the density functions of charge and isospin roots, the
kinetic energy per length has the form $E_k/L=\int k^2\rho(k)dk$,
the total number of down isospins is $M/L=\sum_n
n\int\sigma_n(\lambda)d\lambda$ and the particle density of the
model is $D=N/L=\int \rho(k) dk$. If we consider the energy
arising from the external field $\Omega$ which is the Rabi field
in two-component BEC experiments, the internal energy of the
model is
\begin{equation}
E/L=\int(k^2-\Omega)\rho(k)dk+\sum_n 2n\Omega\int\sigma_n
d\lambda.
\end{equation}
And with the help of the approach first introduced by Yang and
Yang\cite{CNYang69}, the entropy of the present model at finite
temperature is
\begin{eqnarray}
&&{\cal S}/L=\int [(\rho+\rho^h)\ln(\rho+\rho^h)
-\rho\ln\rho-\rho^h\ln\rho^h]dk \\ \nonumber
&&+\sum_n\int[(\sigma_n+\sigma_n^h)\ln(\sigma_n+\sigma_n^h)
-\sigma_n\ln\sigma_n -\sigma_n^h\ln\sigma_n^h]d\lambda.
\end{eqnarray}
The Gibbs free energy of the model then is defined by $F=E-T{\cal
S}-\mu N, $ where $\mu$ is the chemical potential. In order to
obtain the thermal equilibrium, we minimize the free energy with
respect to all the density functions subjects to the constraint
(\ref{eq:DENSITY}). In addition, the total number of particles,
the magnetization are to be keep constant. For this purpose, the
chemical potential $\mu$ and external field $\Omega$ play the
role of Lagrange multipliers.

It is useful to define the energy potential for charge sector and
isospin sector:
\begin{eqnarray}
\kappa(k)&&=e^{\epsilon(k)/T}=\rho^h(k)/\rho(k)
\nonumber \\
\eta_n(\lambda)&&=\sigma_n^h(\lambda)/\sigma_n(\lambda).
\end{eqnarray}
Applying the minimum condition $\delta F=0$ gives rise to a
revised version of Gaudin-Takahashi equations
\begin{eqnarray}
T\ln\kappa=&&\epsilon(k)=k^2-\mu-\Omega-T
K_2(k)*\ln[1+\kappa^{-1}]
 \nonumber \\
&&-T\sum_n K_n(k)*\ln[1+\eta^{-1}]\nonumber \\
\ln\eta_1=&&\frac{1}{4u}{\rm sech}(\pi\lambda/2u)*
  \ln[(1+\kappa^{-1})(1+\eta_2)] \nonumber \\
\ln\eta_n=&&\frac{1}{4u}{\rm sech}(\pi\lambda/2u)*
  \ln[(1+\eta_{n-1})(1+\eta_{n+1})].
\label{thermalequations2}
\end{eqnarray}
And these equations are completed by the asymptotic conditions
\begin{equation}
\lim_{n\rightarrow\infty}[\ln\eta_n/n]=2x \label{eq:ASYCON}
\end{equation}
where $x=\Omega/T$. Eqs. (\ref{thermalequations2}) can be solved
by iteration. Note that eqs. (\ref{eq:DENSITY}) together with eqs.
(\ref{thermalequations2}) completely determine the densities of
charge roots and isospin roots in the state of thermal
equilibrium. The Helmholtz free energy $F=E-T{\cal S}$ is given by
\begin{eqnarray}
F=&&\mu N-\frac{TL}{\pi}\int\ln[1+e^{-\epsilon/T}] dk.
\label{eq:HFE}
\end{eqnarray}

The above approach called TBA is universal for discussing the
thermodynamics of one dimensional integrable model. Once eqs.
(\ref{thermalequations2}) are solved, all thermodynamic
quantities can be obtained from eq. (\ref{eq:HFE}) in principle.

{\bf Magnetic property: zero and high temperature limit:} The
state at zero temperature is the ground state. The Fermi surface
is determined by $\epsilon(k_F)=0$. Since there is no hole under
Fermi surface, we can take the energy potential
$\kappa=\rho^h/\rho$ as zero. As a result, from eqs.
(\ref{thermalequations2}), it is easy to see
$\eta_n\rightarrow\infty$, and  $M=0$, the ``ferromagnetic"
ground state. The first equation of eqs. (\ref{thermalequations2})
becomes
\begin{equation}
\epsilon_0(k)=k^2-\mu-\Omega + K_2(k)*
  \epsilon_0(k)
\end{equation}
which gives the solution of dressed energy, and the ground-state
energy may be given in terms of $\epsilon_0$
\begin{equation}
E_0/L=\frac{1}{2\pi}\int_{-k_F}^{k_F}\epsilon_0(k)dk.
\end{equation}

Consequently, the ground state of 1D SU(2) bosons is an isospin
``ferromagnetic" state, which coincides with the analysis of Li
et al.\cite{YQLi01}. Then the property of the model at $T=0$ is
the same as that of scalar bosons in one dimension which has been
discussed extensively by Lieb and Liniger\cite{EHLieb63}. In the
isospin space, however, the SU(2) symmetry of whole system around
the ground state is broken.

In the high temperature limit $T\rightarrow \infty$ (free
isospins), however we can assume that all functions
$\eta_n(\lambda)$ are independent of $\lambda$. Then eqs.
(\ref{thermalequations2}) can be written as follows,
\begin{eqnarray}
\eta_1^2=&&(1+\eta_2)\nonumber \\
\eta_n^2=&&(1+\eta_{n-1})(1+\eta_{n+1})
\end{eqnarray}
where we have neglected the term $(1+\kappa^{-1})$ in the second
equation of eqs. (\ref{thermalequations2}). The solution of
$\eta_n$ are then constants fixed by the field boundary condition
(\ref{eq:ASYCON}) to be
\begin{equation}
\eta_n=\left[ \frac{\sinh(n+1)x}{\sinh x}\right]^2-1.
\end{equation}

After perform the Fourier transformation on eqs.
(\ref{eq:DENSITY}), we get the solution of the densities of
$\lambda$ n-strings,
\begin{eqnarray}
\sigma_1+\sigma_1^h=&&\frac{1}{4u}{\rm sech}[\pi\lambda/2u]
*[\rho+\sigma_2^h]\nonumber \\
\sigma_n+\sigma_n^h=&&\frac{1}{4u}{\rm sech}[\pi\lambda/2u]*
[\sigma_{n+1}^h+\sigma_{n-1}^h].
\end{eqnarray}
If we assume that $\sigma_n$ and $\sigma^h_n$ are independent of
$\lambda$ or let $c=0$, the total number of down isospins has the
form,
\begin{equation}
\sum_n n\sigma_n=\frac{\rho}{2}-\frac{n_m+1}{2}\sigma_{n_m}e^{n_m
\Omega/T} \label{eq:ML}
\end{equation}
where $n_m$ is maximal length of $\lambda$ strings. In the
absence of Rabi field, we have $M/N=1/2$, the system at high
temperature is a quasi ``paramagnetic" state. If the external
field $\Omega$ is small, expanding eq. (\ref{eq:ML}) for small
filed $x$ and integrating the equation over $\lambda$ space, we
get the magnetization of the model, let $M_m$ be the total number
of isospin rapidities in all $n_m$-strings,
\begin{equation}\label{eq:MAGNETIZATION}
\frac{S_z}{L}=\frac{M_m}{2L}\left(1+\frac{n_m
\Omega}{T}+\frac{n_m^2\Omega^2}{2T^2}+\cdots\right)
\end{equation}
where the first term in the brackets arises from
self-magnetization, while the others are contributed by Rabi
field. Eq. (\ref{eq:MAGNETIZATION}) indicate that the magnetic
property of the model in high temperature regime dominated by
Curie's law $\chi\infty 1/T$.

{\bf Strong coupling limit:} When $\eta\rightarrow \infty$,
$K_n(k)=0$, from eqs. (\ref{thermalequations2}) we have
\begin{equation}
\epsilon=k^2-\Omega-\mu.
\end{equation}

The free energy of the system (\ref{eq:HFE}) at low temperature
now can be solved by step integration,
\begin{equation}
F/L=\mu D-\frac{2}{\pi}
  \Bigl[\frac{1}{3}\mu^{3/2}+\frac{T^2\pi^2}{24\mu^{1/2}}\Bigr]
\end{equation}
where the external field is set to zero.

We can not deduce the specific heat directly from the free energy
obtained above because the chemical potential is a function of
temperature. From eqs. (\ref{eq:DENSITY}), the density of charge
rapidity has the form
\begin{equation}
\rho=\frac{1}{2\pi} \frac{1}{1+e^{(k^2-\mu)/T}}.
\end{equation}
Clearly, at zero temperature, the Fermi surface is just the
square  root of the chemical potential, so we have $\mu_0=\pi^2
D^2$. At low temperature, however, it is determined by
$D=N/L=\int \rho(k) dk$. After integration, we have a time
dependent chemical potential
\begin{equation}
\mu=\mu_0\Bigl[1-\frac{\pi^2 T^2}{24\mu_0^2}\Bigr]^{-2}.
\end{equation}
Then the free energy becomes
\begin{equation}
F/L=\mu_0 D\Bigl[1+\frac{\pi^2T^2}{12\mu_0^2}\Bigr]
  -\frac{2}{3\pi}
  \mu_0^{3/2}\Bigl[1+\frac{\pi^2T^2}{4\mu_0^2}\Bigr].
\end{equation}
Since by thermodynamics ${\cal S}=-\partial F/\partial T$ and
$C_v=T\partial {\cal S}/\partial T$, we find the specific heat at
low temperature is Fermi-liquid like
\begin{equation}
{\cal S}=C_v=\frac{T}{6D}.
\end{equation}
It is the same as the result of one-component case, since for the
strong coupling limit the isospin and the charge are decoupled
completely, the contribution of isospin to the free energy
vanishes.

{\bf Weak coupling limit:} In order to discuss the possibility of
the existence of BEC, we consider the problem in weak coupling
limit $u\rightarrow 0$. And isospin-isospin reaches its maximal
correlation. At low temperature, however, we do not take string
hypothesis for simplicity. Because $\lim_{c\rightarrow
0}K_n(x)=\delta(x)$, together with eqs. (\ref{eq:DENSITY}) and
eqs. (\ref{thermalequations2}), we obtain
\begin{equation}\label{eq:DENSITYOF0}
\rho(k)=\frac{1}{2\pi}\frac{(3e^{\varepsilon_0}-1)(e^{-\varepsilon_0}+1)}
{(3e^{2\varepsilon_0}+1)(1-e^{-\varepsilon_0})}
\end{equation}
where $\varepsilon_0=(k^2-\mu)/T$. The positive definition of
$\rho(k)$ require that the chemical potential is negative. As we
known the density of scalar boson is
$2\pi\rho=1/(1-e^{-\varepsilon_0})$ which prevents the BEC in 1D
and 2D system because of the infrared divergence. However, the
density function (\ref{eq:DENSITYOF0}) still does not resolve this
problem. Consequently, BEC does not happen in this model.

To summarize, we discussed the general thermodynamics of one
dimensional SU(2) bosons with $\delta$-function interaction by
using the strategy of TBA. It was shown that the ground state is
an isospin ``ferromagnetic" state which differs from the ground
state of 1D fermions, while at high temperature, it is
``paramagnetic" state and the magnetic property is dominated by
Curie's law. In strong coupling limit, we obtain the exact
expression of the dependence of chemical potential, entropy and
specific heat on temperature which are Fermi-liquid like, while
in weak coupling limit, we found the infrared divergence of charge
roots density function prevents the existence of BEC.

This work is supported by Trans-Century Training Program
Foundation for the Talents and EYF98 of China Ministry of
Education. SJG thanks D.Yang for helpful discussions.

\end{document}